\documentclass[12pt]{article}

\usepackage{newtxtext,newtxmath}

\usepackage{graphicx}
\graphicspath{ {./Figures/} }
\usepackage[letterpaper,margin=1in]{geometry}

\renewenvironment{abstract}
	{\quotation}
	{\endquotation}

\date{}

\makeatletter
\renewcommand{\fnum@figure}{\textbf{Figure \thefigure}}
\renewcommand{\fnum@table}{\textbf{Table \thetable}}
\makeatother

\usepackage{scicite}

\usepackage{url}

\def\scititle{
	Constraint ratio controls viscosity in shear thickening suspensions
}
\title{\bfseries \boldmath \scititle}

\author{
	Qinghao Mao$^{1,2\ast}$,
	Michael van der Naald$^{3}$,
	Abhinendra Singh$^{4}$,
    Heinrich M. Jaeger$^{1,2}$\and
	\small$^{1}$Department of Physics, University of Chicago, Chicago, IL 60637, USA.\and
	\small$^{2}$James Franck Institute, University of Chicago, Chicago, IL 60637, USA.\and
    \small$^{3}$Department of Molecular and Cellular Biology, Harvard University, Cambridge, MA 02138, USA.\and
    \small$^{4}$Department of Macromolecular Science and Engineering, Case Western Reserve University,\\
    \small Cleveland, OH 44106, USA
    \\
	\small$^\ast$Corresponding author. Email: qinghaomao@uchicago.edu\and
}

\begin{document} 

\maketitle

\begin{abstract} \bfseries \boldmath
The dramatic viscosity increase observed in dense suspensions under shear poses a major challenge in our understanding of how microscopic contact mechanics translate into macroscopic flow resistance. Here, we introduce a constraint-counting model that incorporates friction and dimensionality naturally without additional assumptions and allows for collapsing of rheological data onto a universal master curve. In this model, we borrow ideas from dry granular jamming physics and classify contacts as either locked or non-locked to define a single state variable, the constraint ratio, which measures the average strength of mechanical constraint per particle. By identifying the constraint ratio as the key control parameter, our framework provides a unifying route toward predictive modeling and rational design of shear-thickening materials.
\end{abstract}

\newpage

\noindent
Shear thickening is widely observed in dense suspensions and has many important practical applications\cite{brown2014RPP, morris2020ARFM}. In extreme cases, the viscosity can increase by several orders of magnitude with only a small change in shear rate\cite{barnes1989JOR}, a phenomenon known as discontinuous shear thickening (DST). Over the past two decades, extensive experimental and numerical work has established that frictional interparticle contacts play a central role in this dramatic response to applied shear\cite{Seto2013PRL, wyart2014PRL, Lin2015PRL, peters2016Nature, royer2016PRL, clavaud2025quick}. More recently, analyses using the information of the frictional network have added new insights into how collective contact locking promotes thickening\cite{thomas2020JOR, gameiro2020PRF, vandernaald2024NP, santra2025PRR, aminimajd2025SM, sharma_2026, d_2025}. However, important unresolved issues remain in the theoretical understanding of shear thickening.

While mean-field approaches, most prominently the Wyart-Cates model, reproduce DST qualitatively\cite{wyart2014PRL, mari2014JOR, mari2015PNAS}, quantitative agreement with the rheological flow curves obtained from experiments typically requires substantial parameter fitting. For example, recent studies achieve data collapse only by introducing a stretched exponential function to model the fraction of frictional contacts, together with fitting a nonmonotonic packing-fraction-dependent variable to modulate that fraction\cite{ramaswamy2023JOR, ramaswamy2025PRL}. Such procedures, while effective, highlight the absence of a state variable directly linking microscopic contact statistics to macroscopic viscosity.

A related issue concerns the role of the interparticle friction coefficient. In simulations, pronounced DST is often obtained only for relatively large friction coefficients, commonly $\mu = 1$\cite{Seto2013PRL, mari2015PNAS, vandernaald2024NP, santra2025PRR}, whereas smaller $\mu$ typically yields weakened or continuous thickening responses\cite{mari2014JOR, singh2018JOR, singh2022PRF}. Although this trend is qualitatively understood, there is currently no theoretical relation that quantitatively predicts how the viscosity depends on $\mu$ across different friction regimes. The friction coefficient thus remains an empirical control parameter rather than an integrated element of a predictive theory.

Dimensionality poses an additional conceptual challenge. While simulations indicate broadly similar rheological behavior in two and three dimensions, rigidity-based analyses on the underlying contact network structures, which are effective in 2$D$, encounter fundamental obstacles in 3$D$, where uniquely defined rigid clusters are difficult to identify deterministically\cite{chubynsky2007PRE}. Consequently, no specific dimension-independent geometric or topological structure has been established as the fundamental unit underlying shear thickening.

Taken together, these issues expose persistent gaps between microscopic contact-level descriptions and macroscopic rheology. A unifying method, independent of dimension, that can quantitatively connect viscosity with frictional contact evolution is still missing.

Here, we introduce a constraint-counting approach inspired by studies of frictional granular jamming. Applying this approach to detailed simulations of dense suspensions, we show that a single variable,  which we term the constraint ratio, governs viscosity evolution across different packing fractions, shear rates and stresses. This framework quantitatively collapses rheological curves for all friction coefficients and is applicable to both 2$D$ and 3$D$.

\subsection*{Methods}
The shear-thickening simulations are performed using the well-established Lubricated-Flow Discrete Element Method (LF-DEM) framework~\cite{morris2020ARFM}. The simulation scheme marries near-field hydrodynamic lubrication interactions with a discrete-element-method (DEM) contact model and has been shown to reproduce a wide range of experimentally observed features of shear thickening in dense suspensions~\cite{singh2018JOR, singh2022PRF}. We simulate simple-shear flow of non-Brownian frictional bidisperse spheres (size $a$ and $1.4a$) immersed in a Newtonian fluid using Lees–Edwards periodic boundary conditions~\cite{Mari_2015, singh2018JOR}. 

\begin{figure}
\centering
\includegraphics[width=88mm]{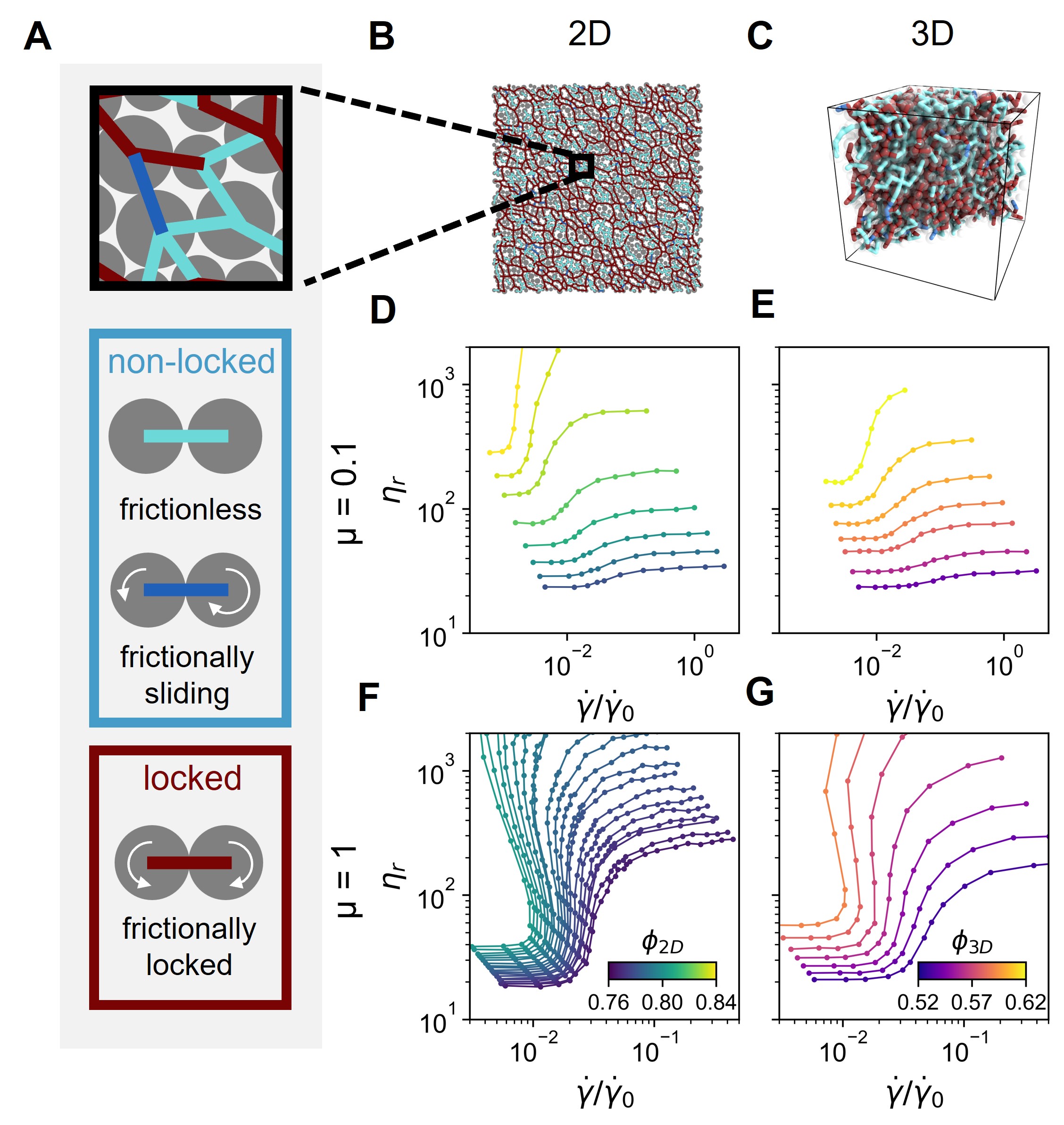}
\caption{{\bf Contact classification and rheological flow curves}. ({\bf A}) Three types of interparticle contacts include (i)frictionless interactions mediated by lubrication forces, (ii) frictional contacts where the Coulomb threshold has been exceeded and sliding occurs, and (iii) frictional contacts below the Coulomb threshold.  These types are grouped into two sets: locked and non-locked. Snapshots of the contact networks obtained from simulations: ({\bf B}) 2$D$, ({\bf C}) 3$D$. Rheological flow curves for different particle packing fractions $\phi$: ({\bf D}) 2$D$, $\mu=0.1$, ({\bf E}) 2$D$, $\mu=1$, ({\bf F}) 3$D$, $\mu=0.1$, ({\bf G}) 3$D$, $\mu=1$. The packing fraction, indicated by the color of the curves, is specified by the scales in F for 2$D$ and in G for 3$D$ systems. Similar rheological behavior, including strong or even discontinuous shear thickening, is obtained if the packing fraction is adjusted appropriately. }
\label{fig_intro}
\end{figure}

Contact interactions are implemented using the approach of Cundall and Strack~\cite{Cundall_1979}. The tangential contact force obeys the Coulomb friction law $\left| \mathbf{F}^{\mathrm C}_t \right|
\le
\mu \,
\left| \mathbf{F}^{\mathrm C}_n \right|$
for compressive normal forces with $\mu$ being the friction coefficient. When the contact between two particles are within the Coulomb threshold, their translational and rotational motions are locked. These locked contacts are shown in red in Fig.~\ref{fig_intro}A. If a contact is sliding, the two motions decouples, and in Fig.~\ref{fig_intro}A it is shown in deep blue.

To incorporate stress-dependent rheology, we employ the critical load model (CLM), in which friction is activated only when the normal contact force exceeds a threshold $F_0$~\cite{Seto2013PRL}. The force $F_0$ sets a characteristic stress scale $\sigma_0 \equiv \frac{F_0}{6\pi a^2}$ which, up to a $\mathcal{O}(1)$ prefactor, corresponds to the crossover from frictionless lubricated particle interactions to direct frictional contact. The frictionless lubricated contacts are shown in light blue in Fig.~\ref{fig_intro}A.

We simulate $N=2000$ particles for 2$D$ simulations (Fig.~\ref{fig_intro}B) and $N=3000$ particles for 3$D$ simulations (Fig.~\ref{fig_intro}C) in a unit cell. Details of the simulation are discussed in Methods.

\subsection*{Rheology curves}

Fig.~\ref{fig_intro}D-G show rheology curves for two friction coefficients ($\mu=0.1$ and $1$) in both 2$D$ and 3$D$. The curves differ substantially between friction coefficients. Systems with higher friction coefficients exhibit stronger shear-thickening behavior, while it becomes much weaker for the $\mu=0.1$ systems. This matches with previous findings\cite{mari2014JOR}, and indicates that the friction coefficient, or equivalently the strength of the constraint imposed by a frictional contact between a pair of particles, has a strong influence on global rheological properties. Despite this importance, the role of friction coefficient has not been carefully examined in many simulations and theoretical studies, which often adopt $\mu=1$ to best match experimental observations of DST. Notably, despite the pronounced differences caused by friction, the rheology curves in 2$D$ and 3$D$ are qualitatively similar if the packing fraction $\phi$ is adjusted, suggesting that the underlying mechanism should be dimension-independent.

\subsection*{Constraint ratio}

To develop a dimension-independent method across different friction coefficient, we adopt Maxwell’s constraint-counting approach and classify contacts into two types: the locked and non-locked, Fig.~\ref{fig_intro}A. The locked (L) contacts are just the frictionally locked ones, whose rotational motion and translational motion are locked. The non-locked (NL) ones consist of the others, the frictionless and frictionally sliding contacts.

We count the number of these two contact types and define $Z_L$ and $Z_{NL}$ as the numbers of locked and non-locked contacts per particle, respectively. 
Using these quantities, we define a mean-field parameter, the constraint ratio $\chi(Z_L, Z_{NL}) \in [0,1]$, with the goal of establishing a one-to-one mapping between $\chi$ and the reduced viscosity $\eta_r$. Here $\eta_r$ is the suspension viscosity normalized by the viscosity of the suspending fluid. We define $\chi=0$ as the dilute limit, where particle motion is unconstrained and $\eta_r$ is minimal, and $\chi=1$ as the jamming limit, where all modes are constrained and $\eta_r$ diverges.

An increase in either $Z_L$ or $Z_{NL}$ increases $\chi$ and therefore enhances viscosity. To first order, we model $\chi$ using a linear relation,
$$
\chi=\alpha Z_{NL} + \beta Z_L.
$$

The coefficients $\alpha$ and $\beta$ can be determined as follows. For a frictionless packing in $d$ dimensions, jamming occurs at $Z_0^J=2d$, implying that each unlocked contact per particle should contribute $\frac{1}{2d}$ toward constraining a particle’s motion. Accordingly, we set $\alpha=\frac{1}{Z_0^J}$ for non-locked contacts. For frictional packings in $d$ dimensions with friction coefficient $\mu$, the jamming contact number $Z_\mu^J$ decreases from $2d$ to $d+1$ as $\mu$ increases from $0$ to infinity\cite{song2008NATURE, wang2010PHYSICAA, silbert2010SM, papanikolaou2013PRL}. For every $Z_\mu^J$ frictional contacts in the system, one particle is frictionally constrained on average. So, $\beta=\frac{1}{Z_\mu^J}$ and it varies from $\frac{1}{2d}$ to $\frac{1}{d+1}$ as a function of $\mu$. With both $\alpha$ and $\beta$ determined, this model has no free parameters and can be tested directly against simulation data.

\subsection*{Collapsing of 2$D$ and 3$D$ viscosity curves}

We first test the model using 2$D$ simulation data. For $d=2$, the theoretical value is $\alpha=1/4$, and $\beta$ goes from $1/4$ to $1/3$, which can be obtained directly from studies of the frictional jamming transition. We use contact numbers at the jamming point from previous work to compute $\beta$\cite{silbert2010SM}. With these numbers, we find that for all friction coefficients the viscosity curves are collapsing on a single curve,  Fig.~\ref{fig_raw_collapse}A. Since the $\alpha$ and $\beta$ values are derived from dry granular simulations, they may differ slightly from those in our suspension simulations. We therefore vary $\alpha$ and $\beta$ and find that the best collapses we can find, Fig.~\ref{fig_collapse_2d}A, are only slightly better. Also see  Fig.~\ref{fig_raw_collapse}.

The $\alpha$ and $\beta$ values for the best collapses are determined by optimizing a loss function, defined as
$L_\mu = L_{\mu,\text{collapse}} + L_{\mu,\text{diverge}},$
for each $\mu$,
where $L_{\mu, \text{collapse}}$ quantifies the quality of curve collapse, and $L_{\mu, \text{diverge}}$ enforces divergence at $\chi=1$, the maximum possible value. Details are provided in the Methods. We sweep $\alpha$ and $\beta$ on a grid, and Fig.~\ref{fig_collapse_2d}B shows the resulting loss landscape. The white dashed lines indicate theoretical predictions from Ref.~\cite{silbert2010SM}, and the gray crosses represent one standard error of the optimal parameters, $\alpha_\mu^0$ and $\beta_\mu^0$. The optimized values agree well with the theory. This suggests that the parameters in our simple constraint ratio model $\chi$ are working very closely at its best, confirming the deep connection between dry granular physics and suspension physics.

\begin{figure}
\centering
\includegraphics[width=88mm]{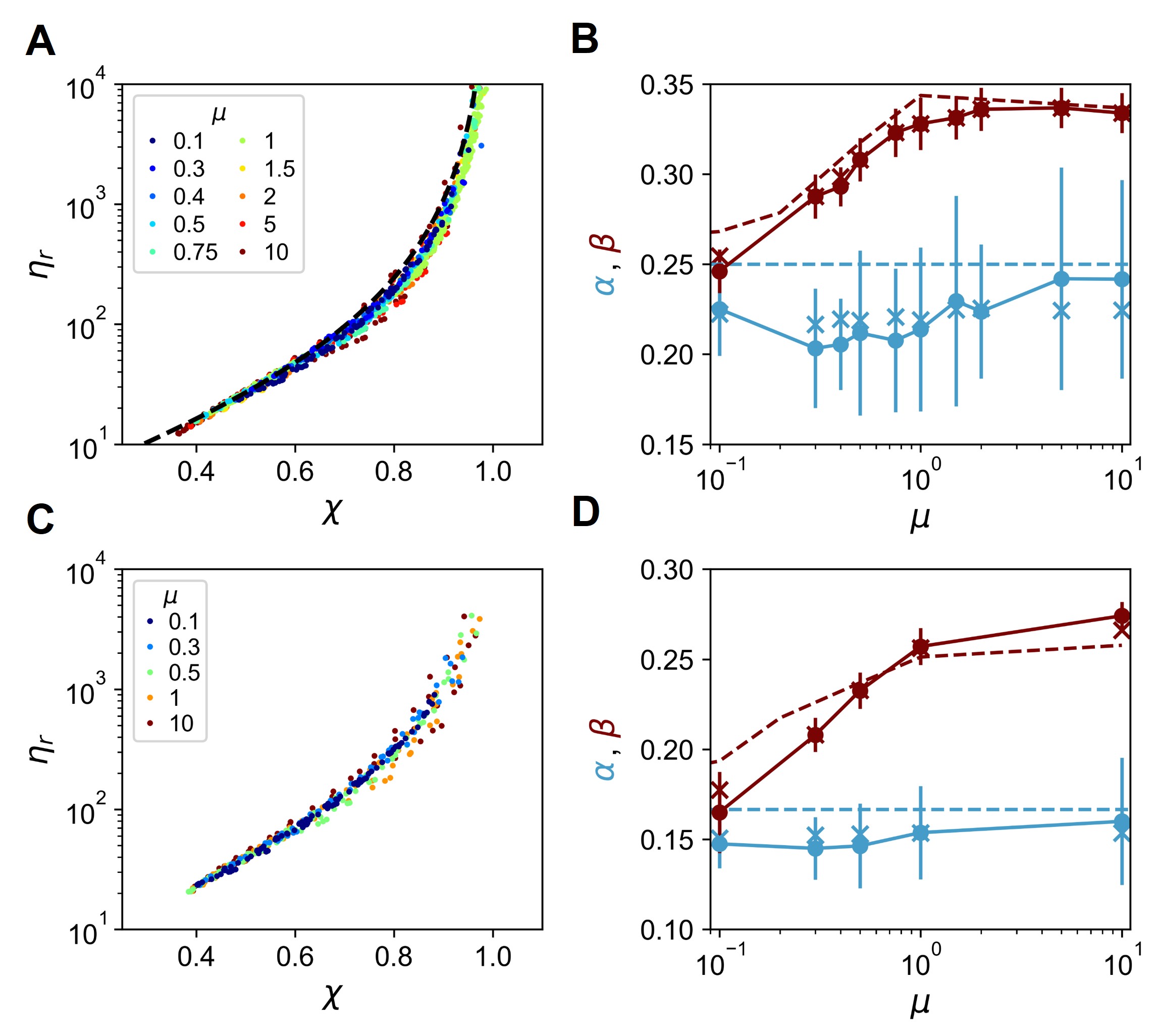}
\caption{{\bf Viscosity master curves in both 2$D$ and 3$D$}. The collapsing of viscosity curves for different $\mu$ in ({\bf A}) 2$D$ and ({\bf C}) 3$D$. Comparison of the best fit values and the theoretical ones in ({\bf B}) 2$D$ and ({\bf D}) 3$D$.}

\label{fig_collapse_2d_3d}
\end{figure}

We further demonstrate that the master curves for different $\mu$ are nearly identical, as shown in Fig.~\ref{fig_collapse_2d_3d}A. The parameters used, $\alpha_\mu$ and $\beta_\mu$, marked as black crosses in Fig.~\ref{fig_collapse_2d}A, are only a tiny variation (around 0.01) away from $\alpha_\mu^0$ and $\beta_\mu^0$. A detailed comparison between fitted and theoretical parameters is shown in Fig.~\ref{fig_collapse_2d_3d}B. Also see details in Methods. This demonstrates that the constraint ratio $\chi$ is a single controlling parameter for viscosity across different friction coefficients and packing fractions.  We notice that the classification of frictionally sliding contacts does not have a strong influence on the collapsing, see Fig.~\ref{fig_raw_collapse}B, which might come from the low proportion of these contacts at high $\mu$, Fig.~\ref{fig_p_frictionally_sliding}, and the closeness of $\alpha$ and $\beta$ values at low $\mu$.

Because the constraint ratio framework is dimension-independent, it naturally extends to 3$D$ systems. Fig.~\ref{fig_collapse_2d_3d}C shows the collapse of $\eta(\chi)$ curves in 3$D$, and Fig.~\ref{fig_collapse_2d_3d}D compares fitted and predicted parameters. These results confirm that the constraint ratio framework applies universally across dimensions, although the function $\eta_r(\chi)$'s shape is slightly different, which might come from the rigidity structure's difference in different dimensions. 

\subsection*{Constraint network structures}

\begin{figure}
\centering
\includegraphics[width=172mm]{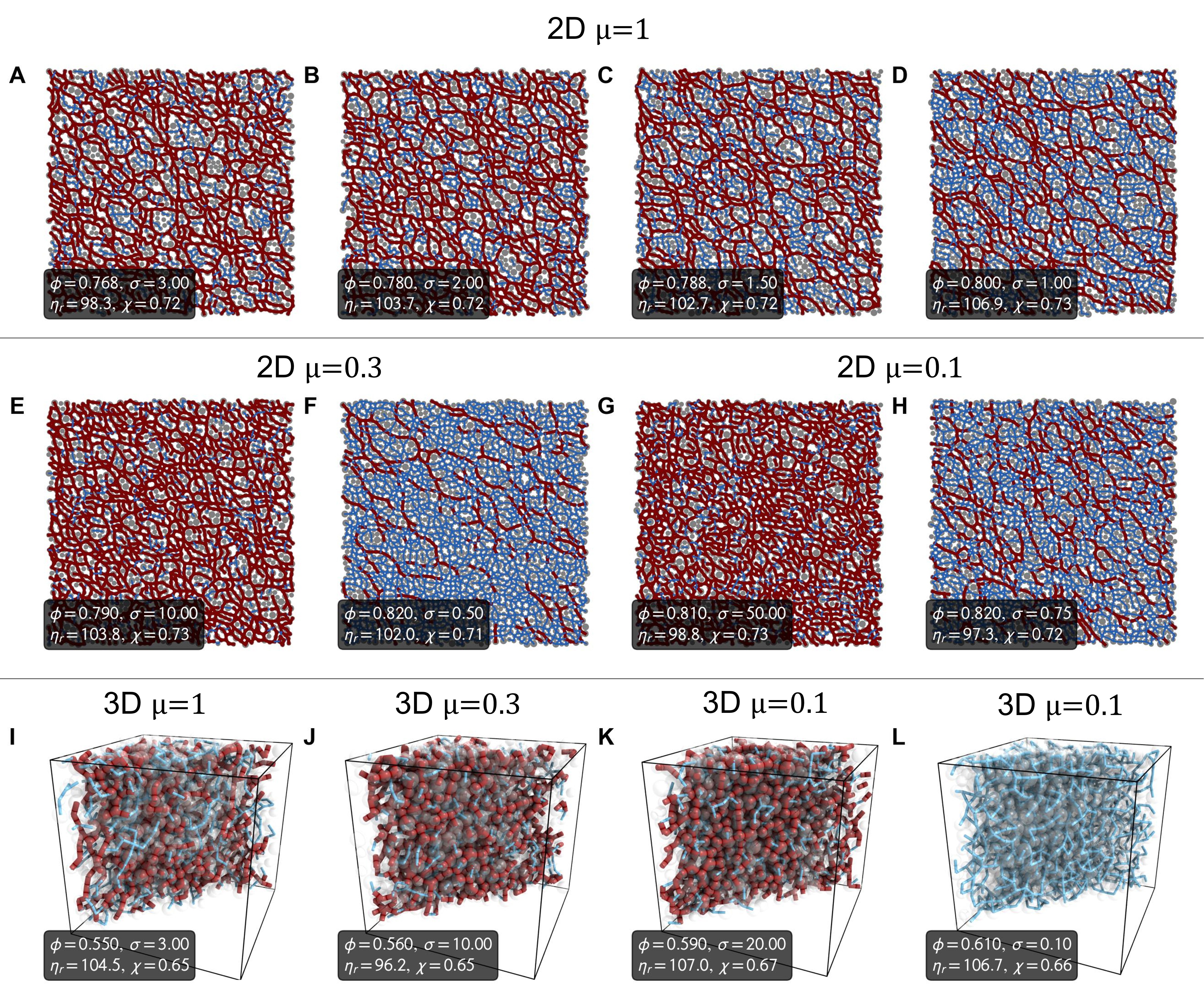}
\caption{{\bf Constraint ratio $\chi$ predicts viscosity for widely differing contact networks.} All systems are chosen with $\chi=0.72\pm0.01 (2$D$), 0.66\pm0.01 (3$D$)$ and $\eta_r=100\pm10$. The locked contacts are shown as thick red bonds and the unlocked ones as thin blue.}
\label{fig_network}
\end{figure}

By labeling contacts with their constraint type, we study the structure with the associated network of constraints. We examine systems with similar $\chi=0.72\pm0.01 (2$D$), 0.66\pm0.01 (3$D$)$ (and thus similar $\eta_r=100\pm10$) but very different $\mu$, $\phi$, and $\sigma$, as shown in Fig.~\ref{fig_network}. Panels A-D show the 2$D$ $\mu=1$ system at decreasing packing fraction and increasing shear stress. Locked contacts are shown as thick red bonds, and non-locked contacts as thin blue bonds. At high packing fraction, the system consists of a dense locked-contact network decorated with local non-locked contacts. At low packing fraction, only sparse locked structures remain, embedded in a dense non-locked network. Notably, the non-locked network in panel D is denser than the locked network in panel A. This follows directly from $\chi=\alpha Z_{NL} + \beta Z_L$: since $\alpha<\beta$, achieving the same $\chi$ requires more non-locked contacts than locked contacts.

Similar behavior is observed for $\mu=0.3$ (panels E and F) and $\mu=0.1$ (panels G and H). The density of the non-locked network is nearly identical across friction coefficients (see panels D, F, and H), while the locked network becomes denser as $\mu$ decreases (see panels A, E, and G). This again reflects the requirement of maintaining constant $\chi$: while $\alpha$ is independent of $\mu$, $\beta$ decreases with decreasing $\mu$, so systems with lower friction coefficients require more locked contacts to reach the same constraint ratio.

Extending this approach to 3$D$, panels I-L, we again see that, for different friction and packing fraction and shear stress, the networks look very different. While no obvious structures can be identified in these networks, our mean-field constraint ratio predicts the viscosity  well.

These findings show that the viscosity is governed by the average mechanical constraint per particle, implying that a mean-field description is sufficient for quantitative prediction. 

\subsection*{Wyart-Cates, Rigidity and DST}

Having identified a universal $\eta(\chi)$ master curve, we now examine its properties using the 2$D$ $\mu=1$ system, for which many other simulational works are based on, and for which extensive rigidity-network studies exist\cite{vandernaald2024NP}.

\begin{figure}
\centering
\includegraphics[width=88mm]{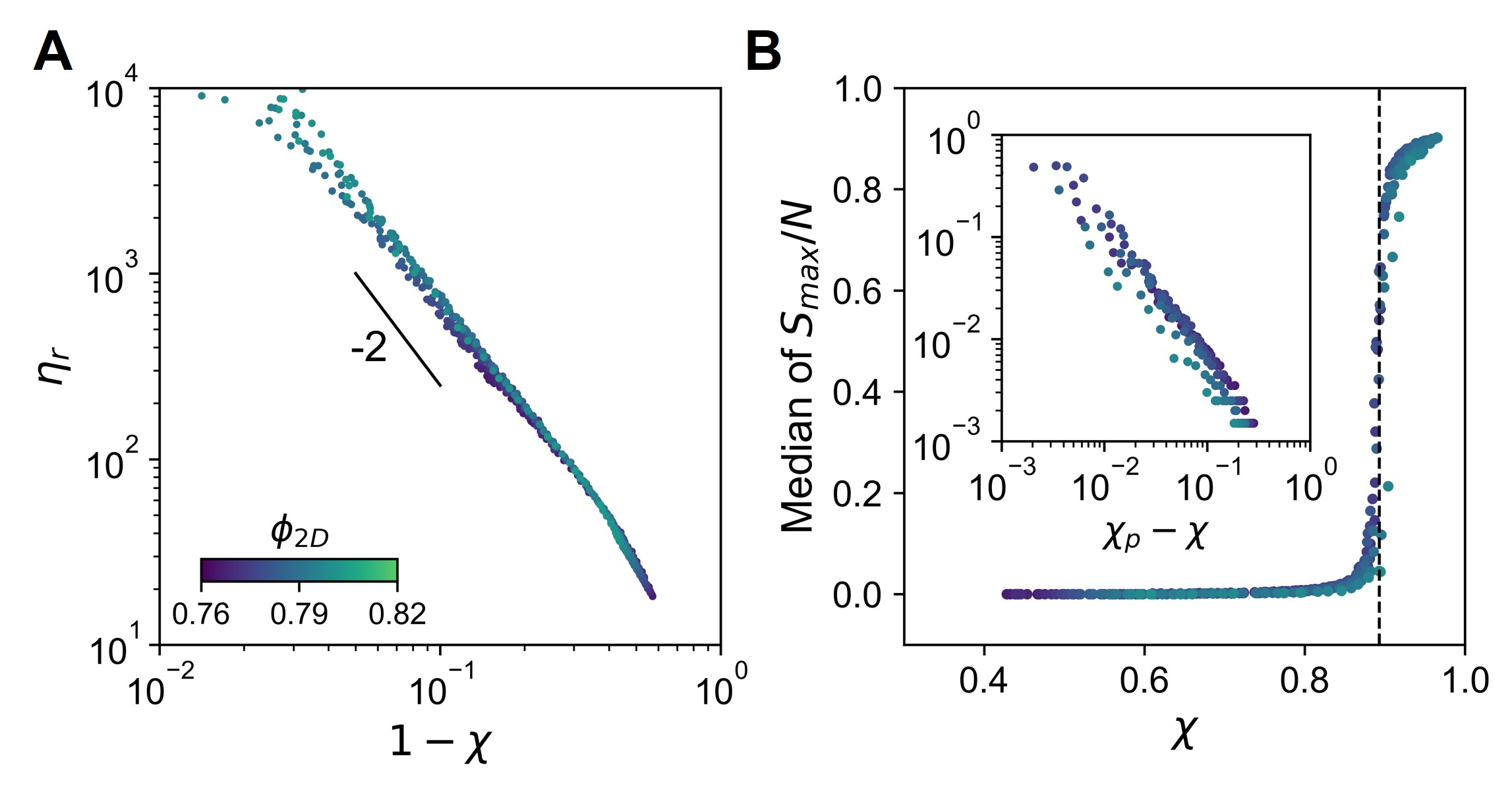}
\caption{{\bf Connect constraint ratio $\chi$ with rigid percolation and jamming}. ({\bf A}) Power-law-like relationship between $\eta_r$ and $1-\chi$, as the system approaching the jamming transition. ({\bf B}) The median size of rigid clusters as a function of $\chi$, suggesting a percolation transition happening at $\chi_p$. Inset: log-log plot of $S_{max}/N$ with $\chi_p-\chi$.}

\label{fig_power_law_rigid}
\end{figure}

As shown in Fig.~\ref{fig_collapse_2d_3d}A, viscosity diverges at $\chi=1$. We therefore plot $\eta_r$ as a function of $1-\chi$ in Fig.~\ref{fig_power_law_rigid}A, revealing a power-law-like behavior with slope $-2$. If we take the simplification of $\beta=\alpha=\frac{1}{Z_C}$, where $Z_C$ is the jamming contact number, we have $1-\chi=\frac{1}{Z_C}(Z_C-(Z_{NL}+Z_L))$. Together with the well-known scaling $\eta_r\propto (\phi_J-\phi)^{-2}$, where $\phi_J$ is the jamming packing fraction, we therefore find that the number of soft modes $\delta_z=Z_C-(Z_{NL}+Z_L)$ scales linearly with $\phi_J-\phi$, which is equivalent to the equations derived in Ref.~\cite{wyart2014PRL}. Thus, we see that our $\eta(\chi)$ model matches the Wyart-Cates model if we consider $\alpha=\beta$.

\begin{figure}
\centering
\includegraphics[width=88mm]{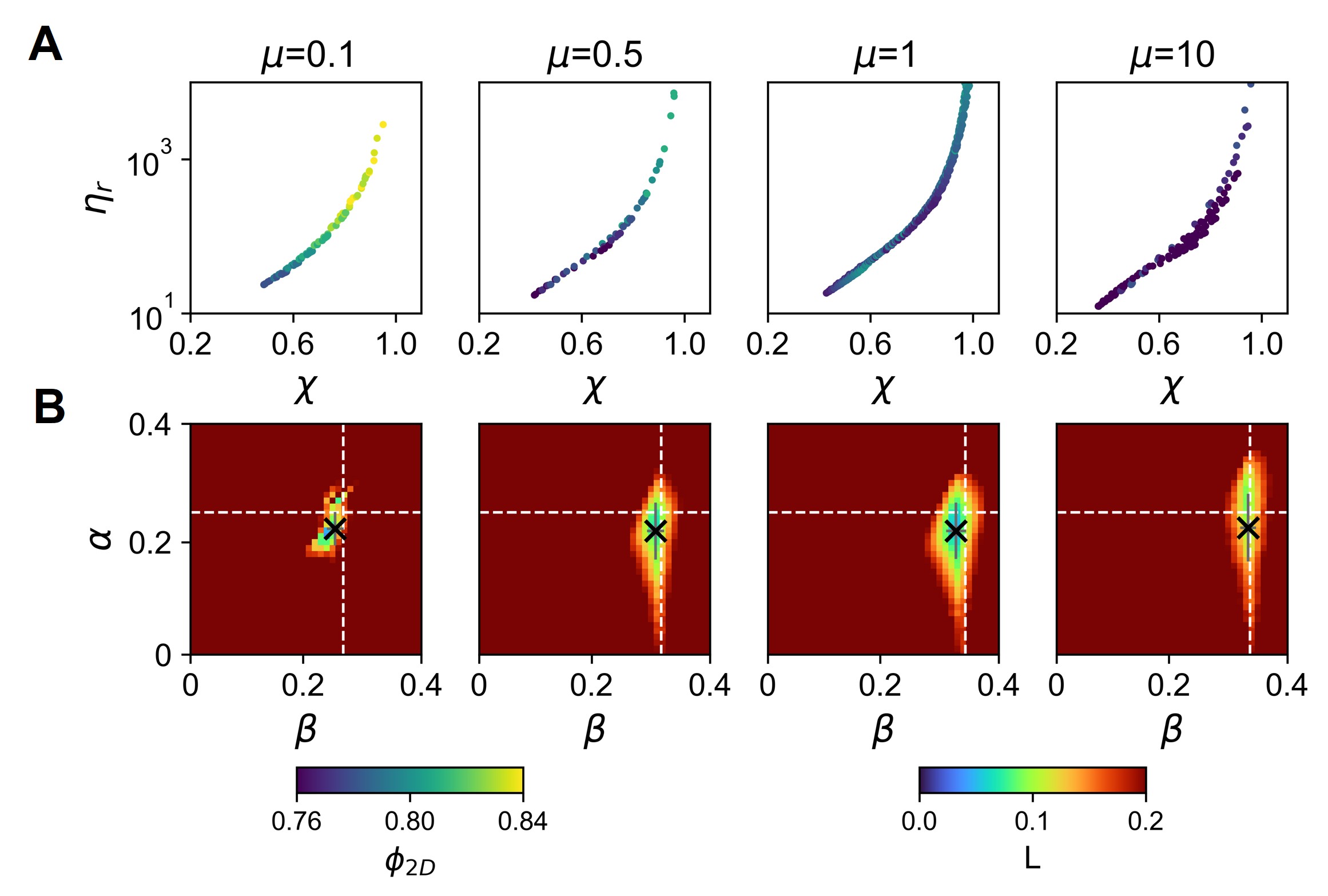}
\caption{{\bf Viscosity master curves in 2$D$}. ({\bf A}) Collapsing of viscosity curves for different $\phi$ and $\mu$ using best $\alpha$ and $\beta$ values. ({\bf B}) The loss function $L$ as a function of $\alpha$ and $\beta$ for different $\mu$s. Dashed white line: theoretical value. Errorbars: 1 standard error of the best fitting parameters. }

\label{fig_collapse_2d}
\end{figure}

Notice that both $Z_C$ and $\phi_J$ are functions of $\sigma$, and they can also depend on $\phi$, so a fitted activation function $f$ is required to model them in Ref.~\cite{wyart2014PRL}. Here, we circumvent this by directly considering $Z_L$ and $Z_{NL}$. As shown in Fig.~\ref{fig_activation_function}, while at high $\mu$, the activation function is roughly independent of $\phi$, in the low friction region, $\phi$-dependency starts to kick in. Such $\phi$-dependence might cause the difficulty in collapsing experimental curves in Ref.~\cite{ramaswamy2023JOR}. In spite of the $\phi$-dependence of $f$, our model still collapses the curves in a robust way, Fig.~\ref{fig_collapse_2d}B, confirming that it builds a universal relationship between local contacts and global rheological behaviors.

To study the rigidity of the system and its connection with $\chi$, we examine the median size of the largest rigid cluster, $S_{max}$, as a function of $\chi$, shown in Fig.~\ref{fig_power_law_rigid}B. A rigid cluster is defined as a cluster of particles whose number of constraints is larger or equal to the number of its degrees of freedom, and it is identified using the pebble-game as discussed in Ref.~\cite{vandernaald2024NP}. A sharp increase occurs near $\chi_p \approx 0.9$. When plotted against $\chi_p-\chi$, a power-law behavior emerges, indicating a percolation transition as the fraction of constrained particles approaches $\chi_p$. This explains the turning point in the $\eta(\chi)$ curve. Notice that the rigidity percolation happens before $\chi=1$, i.e., not all particles have to be constrained for the rigidity cluster to span the whole system, and viscosity can still increase after percolation. Therefore, this $\chi_p \approx 0.9$ threshold may inspire further research on the formation of rigidity and analysis of structures.

Importantly, we note that the $\eta(\chi)$ model accurately captures viscosity across the whole simulationed range, both below and above DST. To clarify the origin of shear thickening and how DST relates to features in $\chi$, we decompose the slope of the theological flow curves by writing $\frac{d\log\eta_r}{d\log\sigma}=\frac{d\log\eta_r}{d\chi}\times\frac{d\chi}{d\log\sigma}$. This is shown in Fig.~\ref{fig_zn_zl_dst}D-F. At the onset of DST, $\frac{d\log\eta_r}{d\log\sigma}=1$, $\frac{d\log\eta_r}{d\chi}$ is almost constant across different $\phi$, and $\frac{d\chi}{d\log\sigma}$ is close to its maximum. This indicates that the sharp increase in viscosity at DST primarily originates from the rapid growth of the constraint ratio with stress, rather than from any special feature in the functional dependence of $\eta_r$ on $\chi$. While DST appears as a distinct point when shear rate or stress is chosen as the control parameter, $\eta_r$ is a smooth function of $\chi$, with no transitions observed near DST. In this sense, DST is not a distinct state of the system, but a manifestation of rapid constraint evolution within the $\eta_r(\chi)$ framework.

\section*{Conclusion}

We  introduced a constraint ratio model that provides a mean-field description of how local locked and non-locked contacts contribute to the global viscosity increase in shear-thickening suspensions, including the discontinuous regime. The model is  grounded in the theory of frictional granular jamming and does not require fitting parameters. Despite the simplicity in the definition of the constraint ratio $\chi$, it successfully collapses viscosity curves in both 2$D$ and 3$D$ across all friction coefficients, demonstrating strong predictive power, therefore offers guidance for the measurement, design, and optimization of shear-thickening properties in practical applications.

While we find more elaborate network characterizations are not required to reproduce the global rheological response, we believe network-based structural analysis may benefit from explicitly incorporating constraint strength into analysis to provide valuable insights into local dynamics of shear thickening suspensions.

\clearpage 

%
\bibliography{ref} 
\bibliographystyle{sciencemag}

%
%
%
%
%
%


\section*{Acknowledgments}
We thank Aaron Galper and Meera Ramaswamy for insightful discussions. 
\paragraph*{Funding:}
This research was supported by the Army Research Office through W911NF-25-2-0156 and W911NF-21-2-0146.  The research also benefited from computational resources and services supported by the Research Computing Center at the University of Chicago.
\paragraph*{Author contributions:}
Q.M. conceived the study. A.S. performed the simulations. All authors contributed to the analysis of the results and to the writing of the manuscript.
\paragraph*{Competing interests:}
The authors declare no competing interests.
\paragraph*{Data and materials availability:}
Data, code, and any additional materials supporting the findings of this study are available from the corresponding author upon reasonable request.


\subsection*{Supplementary materials}
Materials and Methods\\
Figs. S1 to S4\\


\newpage


\renewcommand{\thefigure}{S\arabic{figure}}
\renewcommand{\thetable}{S\arabic{table}}
\renewcommand{\theequation}{S\arabic{equation}}
\renewcommand{\thepage}{S\arabic{page}}
\setcounter{figure}{0}
\setcounter{table}{0}
\setcounter{equation}{0}
\setcounter{page}{1} 


\begin{center}
\section*{Supplementary Materials for\\ \scititle}


	Qinghao Mao$^{1,2\ast}$,
	Michael van der Naald$^{3}$,
	Abhinendra Singh$^{4}$,
    Heinrich M. Jaeger$^{1,2}$\and
    
	\small$^{1}$Department of Physics, University of Chicago, Chicago, IL 60637, USA.\and
    
	\small$^{2}$James Franck Institute, University of Chicago, Chicago, IL 60637, USA.\and
    
    \small$^{3}$Department of Molecular and Cellular Biology, Harvard University, Cambridge, MA 02138, USA.\and
    
    \small$^{4}$Department of Macromolecular Science and Engineering, Case Western Reserve University,\\
    \small Cleveland, OH 44106, USA
    \\
	\small$^\ast$Corresponding author. Email: qinghaomao@uchicago.edu\and

\end{center}

\subsubsection*{This PDF file includes:}
Materials and Methods\\
Figures S1 to S4\\

\newpage


\subsection*{Materials and Methods}

\subsubsection*{Simulation details}
Because inertia is negligible at the particle scale for the systems considered here, the dynamics in our simulation is treated in the overdamped limit. The equation of motion reduces to a force balance on each particle
\begin{equation}
\mathbf{0}
=
\mathbf{F}^{\mathrm H}(\mathbf{r},\mathbf{u})
+
\mathbf{F}^{\mathrm C}(\mathbf{r}),
\label{eq:force_balance}
\end{equation}
where $\mathbf{r}$ and $\mathbf{u}=\dot{\mathbf{r}}$ denote the positions and velocities of the particles, respectively. The corresponding torque balance also applies. 

Hydrodynamic force on every particle is the sum of one-body Stokes drag due to motion of the particle relative to the surround fluid and two-body lubrication force and is given as:
\begin{equation}
 \mathbf{F}^{\mathrm H}(\mathbf{r},\mathbf{u}) =   -\mathbf{R}_{FU}(\mathbf{r})\cdot (\mathbf{u}-\mathbf{u}^{\infty}) + \dot{\gamma}\mathbf{R}_{FE}(\mathbf{r}): \mathbf{E}_{\infty}~.
\end{equation}
The resistance matrices $\mathbf{R}_{FU}$ and $\mathbf{R}_{FE}$ include the squeeze, shear, and pump modes of pairwise lubrication interactions, as well as single-body Stokes drag~\cite{mari2014JOR, Mari_2015}. The leading resistance diverges as the inverse surface separation $h$ between the particles. To allow contact formation, the lubrication singularity is regularized below a cutoff $h_{\min}=10^{-3}a$ with $h$ being the surface separation between particles~\cite{mari2014JOR}. 

Particle contacts are modeled by linear normal $k_n$ and tangential $k_t$ springs, a simple model widely used in granular physics ~\cite{Cundall_1979}. Tangential and normal contact forces between particles satisfy Coulomb's friction law $\left| \mathbf{F}^{\mathrm C}_t \right|
\le
\mu \,
\left| \mathbf{F}^{\mathrm C}_n \right|$, where $\mu$ being the friction coefficient.

The equation of motion is solved under the constraint of imposed simple shear flow at constant shear stress $\sigma$. At any given time, the total shear stress in the suspension is the sum of the hydrodynamic and contact contributions, 
\begin{equation}
\sigma 
= 
\eta_0 \dot{\gamma}(1+\frac{5}{2}\phi)
+ \sigma^{\mathrm H}
+ \sigma^{\mathrm C},
\label{eq:total_stress}
\end{equation}
where $\eta_0$ is the suspending fluid viscosity and $V$ is the volume of the simulation box. The stress contributions are calculated as: $\sigma^{\mathrm H} = \frac{\dot{\gamma}}{V}
\left\{
(\mathbf{R}_{SE}-\mathbf{R}_{SU} \cdot \mathbf{R}_{FU}^{-1} \cdot \mathbf{R}_{FE}) : \mathbf{E}_{\infty}
\right\}_{xy}$, and $\sigma^{\mathrm C} = \frac{1}{V}
\left\{\mathbf{r}
\mathbf{F}{^{\mathrm C}} - \mathbf{R}_{SU} \cdot \mathbf{R}_{FU}^{-1} \cdot \mathbf{F}{^{\mathrm C}}
\right\}_{xy}$, $\mathbf{R}_{SU}$ and $\mathbf{R}_{SE}$ are position-dependent resistance matrices that determine the lubrication stress arising from particle velocities and deformation resistance, respectively~\cite{Mari_2015, mari2014JOR}; $\mathbf{E}_{\infty}$ is the imposed rate-of-strain tensor; $:$ denotes a double-dot product; and $\{\cdot\}_{xy}$ extracts the $xy$ component of the enclosed tensor.

Imposing a fixed shear stress $\sigma$ leads to the shear rate $\dot{\gamma}(t)$, which is time-dependent, and is used to calculate the apparent viscosity $\eta = \frac{\sigma}{\dot{\gamma}}$. After discarding the start-up transient (which typically persists for $\mathcal{O}(1)$ strain units), we report the relative viscosity $\eta_r \equiv \frac{\sigma}{\eta_0 \langle \dot{\gamma} \rangle},$ where the angle brackets denote a time average taken over the steady-state regime.

\subsubsection*{The Loss function to determine best fit for $\alpha$ and $\beta$}

For each friction coefficient $\mu$ we define a loss
\begin{equation}
L_\mu = L_{\mu,\mathrm{collapse}} + L_{\mu,\mathrm{diverge}},
\end{equation}
where $L_{\mu,\mathrm{collapse}}$ quantifies the smoothness of the $\log_{10}\eta_r(\chi)$ curve (smaller is better) and $L_{\mu,\mathrm{diverge}}$ enforces that the curve diverges at the physically expected point $\chi\approx 1$.

The collapse term is computed by sorting data by $\chi$ and taking the mean absolute jump of the sorted log-viscosity,
\begin{equation}
L_{\mu,\mathrm{collapse}} \;=\; \frac{1}{N_\chi-1}\sum_{k=1}^{N_\chi-1}\big|\Delta\log_{10}\eta_r(\chi_k)\big|,
\end{equation}
which penalizes residual discontinuities after collapse, and it will approach zero when the curve is smoothed enough.

Scale $\alpha$ and $\beta$ by a same factor and $L_{\mu,\mathrm{collapse}}$ does not change. To ensure $\eta_r$ divergence at $\chi=1$, we define a target log-viscosity $\log_{10}\eta_{\max}$ and fit the high-$\chi$ tail with a linear extrapolation
\begin{equation}
\chi(\log_{10}\eta_r)=a\,\log_{10}\eta_r + b,
\end{equation}
evaluated at $\log_{10}\eta_{\max}$; the divergence term is then
\begin{equation}
L_{\mu,\mathrm{diverge}} \;=\; \big|\chi(\log_{10}\eta_{\max}) - 1\big|,
\end{equation}
where $\eta_\text{max}$ is an upper limit for all simulations, which we used $3\cdot10^{4}$ for 2$D$ and $1\cdot10^{4}$ for 3$D$.

We first sweep $(\alpha,\beta)$ on a regular grid and compute $L_\mu$ for each pair to obtain a loss landscape (Fig.~\ref{fig_collapse_2d}A). The per-$\mu$ best fits $(\alpha_\mu^0,\beta_\mu^0)$ and their one-standard-error intervals are extracted from the weighted second moments of the score distribution and are shown as gray crosses in Fig.~\ref{fig_collapse_2d}A.

To further improve cross-$\mu$ consistency, we run a second-layer optimization that simultaneously adjusts per-$\mu$ vectors $\boldsymbol{\alpha}=(\alpha_i)$ and $\boldsymbol{\beta}=(\beta_i)$ to minimize
\begin{equation}
\mathcal{L}(\boldsymbol{\alpha},\boldsymbol{\beta})
=
\sum_i\big[(\alpha_i-\alpha^0_i)^2+(\beta_i-\beta^0_i)^2\big]
\;+\;
L_{\mathrm{smoothness}}(\boldsymbol{\alpha},\boldsymbol{\beta}),
\end{equation}
where $L_{\mathrm{smoothness}}$ is the combined collapse smoothness computed by pooling and sorting the $\chi$–$\log_{10}\eta_r$ data across all $\mu$ and taking the mean absolute jump (same definition as $L_{\mu,\mathrm{collapse}}$ but on the pooled curve). We perform this optimization and the final optimized parameters are marked as black crosses in Fig.~\ref{fig_collapse_2d}A.

\newpage


\begin{figure}
\centering
\includegraphics[width=88mm]{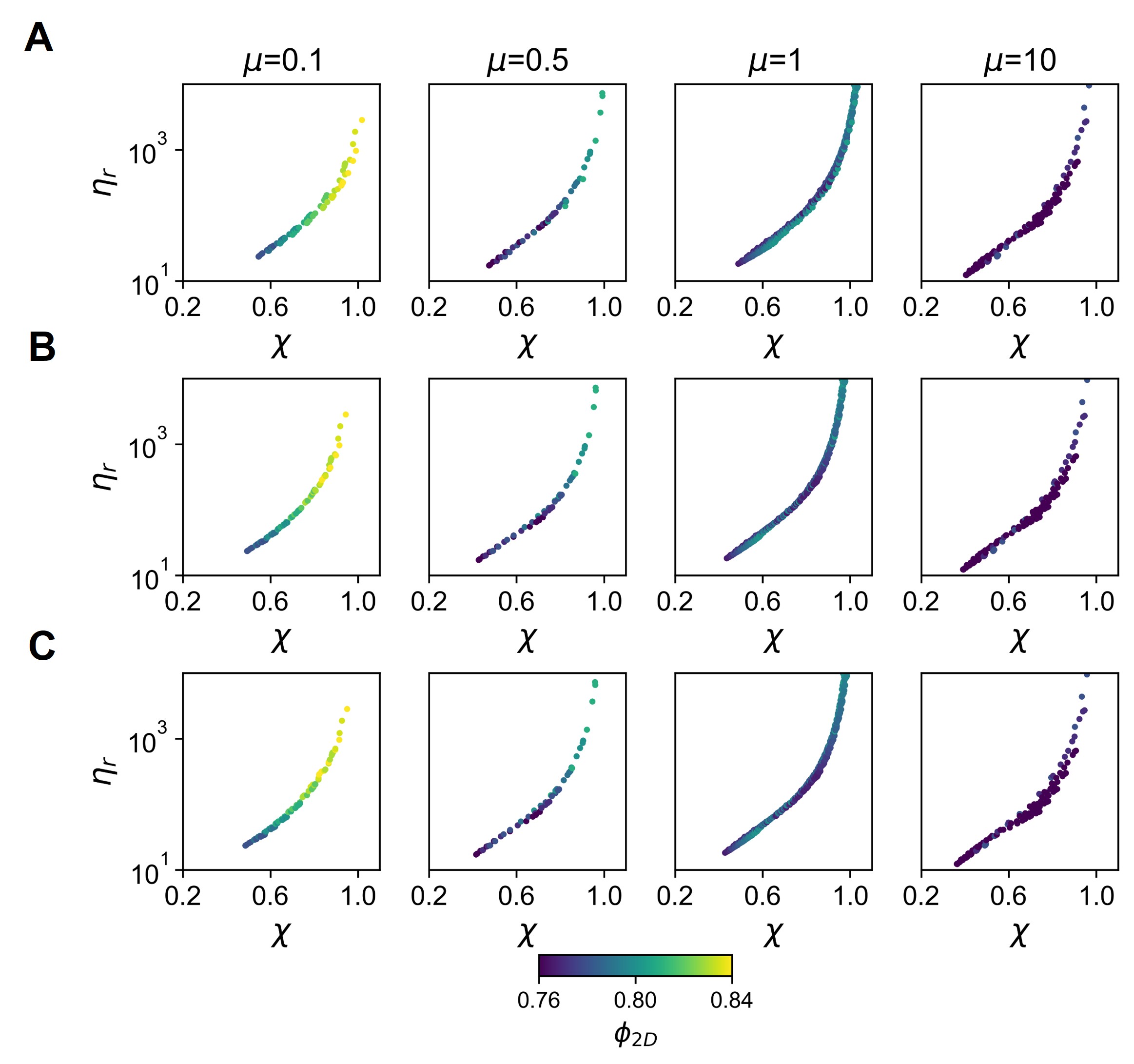}
\caption{{\bf Collapse of viscosity curves in 2$D$ with different methods}. ({\bf A}) Directly using $\alpha$ and $\beta$ values taken from Ref.~\cite{silbert2010SM}. ({\bf B}) Classifing frictionally sliding contacts as locked instead of unlocked, still using the theoretical $\alpha$ and $\beta$ values as in (A). ({\bf C}) Using the best $\alpha$ and $\beta$ values that minimize the loss function. This panel is the same as Fig.~\ref{fig_collapse_2d}A. All three methods give very similar collapses.}
\label{fig_raw_collapse}
\end{figure}

\begin{figure}
\centering
\includegraphics[width=88mm]{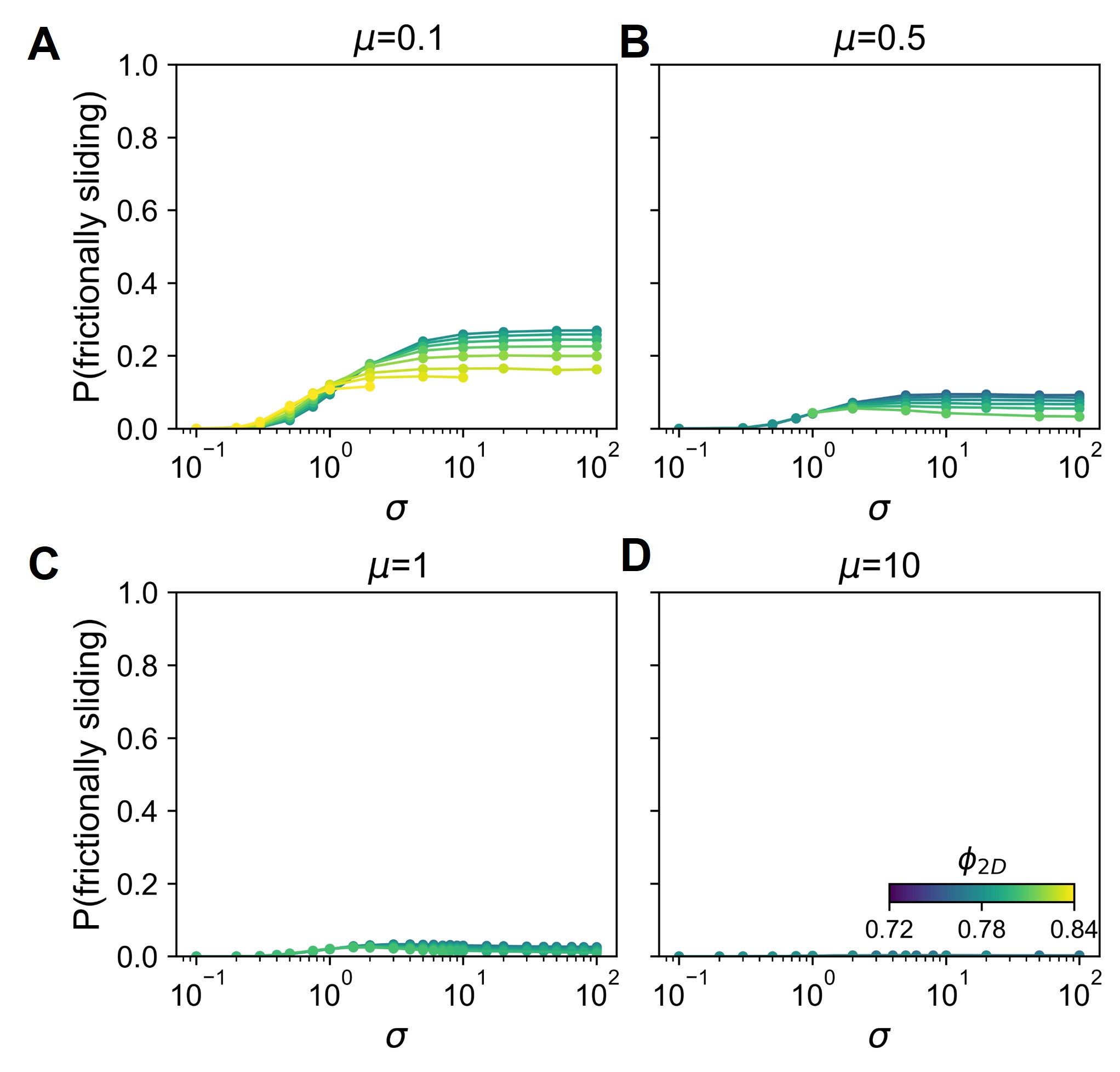}
\caption{{\bf Fraction of frictionally sliding contacts with different $\mu$ in 2$D$.} This fraction is small across all $\mu$ and almost zero at high $\mu$s.}
\label{fig_p_frictionally_sliding}
\end{figure}

\begin{figure}
\centering
\includegraphics[width=88mm]{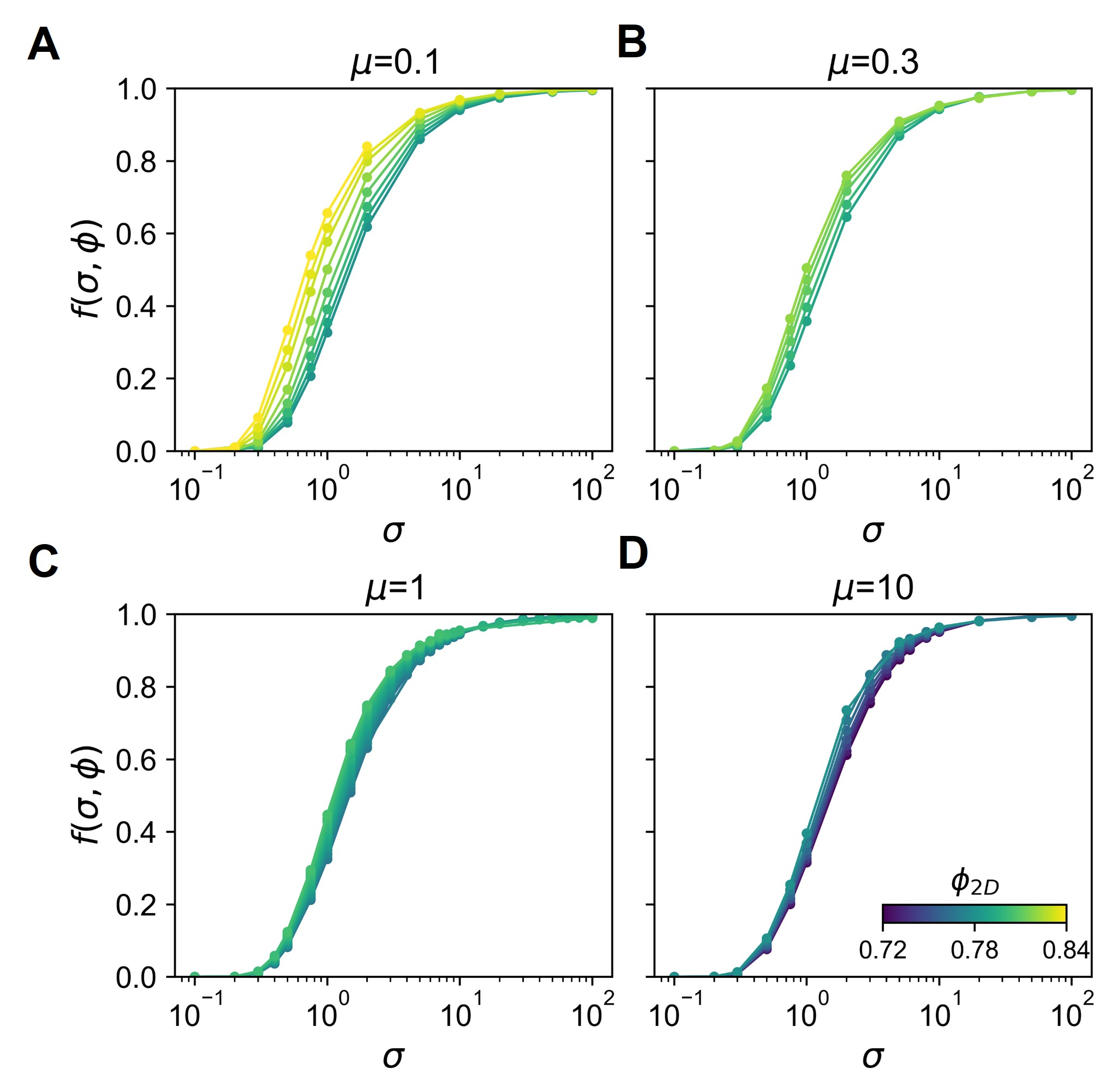}
\caption{{\bf Activation function $f=\frac{Z_L}{Z_{NL}+Z_L}$}. $f$ is almost independent of $\phi$ for $\mu \geq1$, but it varies with $\phi$ for $\mu < 1$.}
\label{fig_activation_function}
\end{figure}

\begin{figure}
\centering
\includegraphics[width=172mm]{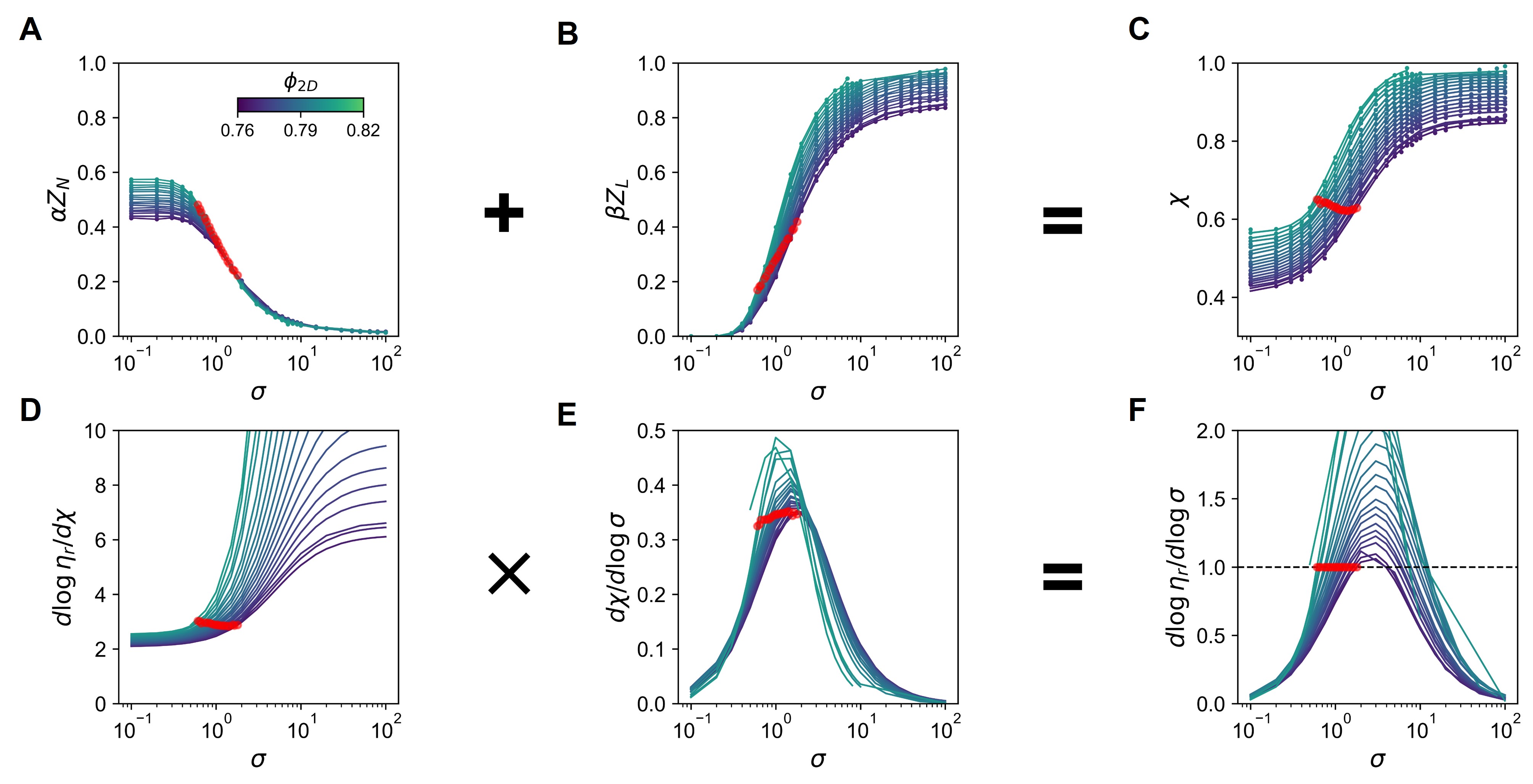}
\caption{{\bf Analysis at DST}. The onset of DST, $\frac{d\log\eta_r}{d\log\sigma}=1$, is marked as a red point for each iso-$\phi$ curve. ({\bf A-C}) $\alpha Z_{NL}$, $\beta Z_L$, and $\chi=\alpha Z_{NL}+\beta Z_L$ as a function of $\sigma$. ({\bf D-F}), $\frac{d\log\eta_r}{d\chi}$, $\frac{d\chi}{d\log\sigma}$ and $\frac{d\log\eta_r}{d\log\sigma}$ as a function of $\sigma$. All data points are from 2$D$ $\mu=1$ simulations.}

\label{fig_zn_zl_dst}
\end{figure}

\end{document}